%% file: Rose-Trial_and_error.tex
\documentclass[10pt,conference]{IEEEtran} 
\usepackage{graphicx}
\usepackage{balance} 
\usepackage{amssymb}
\usepackage{amscd}
\usepackage{dsfont}
\usepackage{amsmath}
\usepackage{rotating}
\usepackage{amsfonts}
\usepackage{url}
\usepackage{color}

\newtheorem{theorem}{Theorem}

\newcounter{def_count}
\newcounter{def_count_cor}
\newtheorem{definition}[def_count]{Definition}

\newtheorem{corollary}[def_count_cor]{Corollary}

\newcommand{\bs}{\boldsymbol}

\newcommand{\defn}{\stackrel{\triangle}{=} }

\newcommand{\bak}{\bar a_k}

\newcommand{\discontent}{\textit{discontent }}
\newcommand{\content}{\textit{content }}
\newcommand{\watchful}{\textit{watchful }}
\newcommand{\hopeful}{\textit{hopeful }}

\graphicspath{{FIGURE/}}

\newcommand{\executeiffilenewer}[3]{%
\ifnum\pdfstrcmp{\pdffilemoddate{#1}}%
{\pdffilemoddate{#2}}>0%
{\immediate\write18{#3}}\fi%
}

\newcommand{\ind}[1]{\mathds{1}_{\left\lbrace #1 \right\rbrace}}

\begin{document}\fontsize{9.3}{11.5}
\title{Distributed Power Allocation with SINR Constraints Using Trial and Error Learning}

\author{
\IEEEauthorblockN{Luca Rose}
\IEEEauthorblockA{Thales Communications,\\
France.\\
 luca.rose@thalesgroup.com}
\and
\IEEEauthorblockN{Samir~M. Perlaza, M\'{e}rouane Debbah}
\IEEEauthorblockA{Alcatel - Lucent Chair in Flexible Radio\\
Supelec, France.\\
(samir.perlaza, merouane.debbah)@supelec.fr}
\and
\IEEEauthorblockN{Christophe J. Le Martret}
\IEEEauthorblockA{Thales Communications,\\
France.\\
christophe.le\_martret@thalesgroup.com}
\and
}
\maketitle


\begin{abstract}
\boldmath 
In this paper, we address the problem of global transmit power minimization in a self-configuring network where radio devices are subject to operate at a minimum signal to interference plus noise ratio (SINR) level. We model the network as a parallel Gaussian interference channel and we introduce a fully decentralized algorithm (based on trial and error) able to statistically achieve a configuration where the performance demands are met. Contrary to existing solutions, our algorithm requires only local information and can learn stable and efficient working points by using only one bit feedback. We model the network under two different game theoretical frameworks: normal form and satisfaction form. We show that the converging points correspond to equilibrium points, namely Nash and satisfaction equilibrium. Similarly, we provide sufficient conditions for the algorithm to converge in both formulations. Moreover, we provide analytical results to estimate the algorithm's performance, as a function of the network parameters. Finally, numerical results are provided to validate our theoretical conclusions.
\end{abstract}

\section{Introduction}\label{SecIntroduction}

In this paper, we consider a network where several transmitter-receiver pairs communicate through a common bandwidth divided into several orthogonal sub-bands, thus, subject to mutual interference. All devices must guarantee certain quality of service (QoS), expressed in terms of signal to interference plus noise ratio (SINR).
The behaviour of the devices is designed for achieving a stable network operating point (equilibrium) where the maximum number of communicating pairs satisfy their QoS with the minimum global power consumption. 
Network operating point must be achieved by the radio devices in a fully decentralized way by selecting their power allocation policy, i.e., selecting a sub-band and a power level for each transmission. In this scenario, all communications take place in absence of a centralized controller and neither cooperation nor exchange of information between different pairs are considered. 
%
For instance, this scenario may model the case of tactical radios and \textit{ad hoc} networks, where, in order to set power and channel, current solutions require a certain level of cooperation with exchange of information between the devices or a manual setting.

\noindent
The closest works to ours are \cite{Iiduka-2010}, \cite{Altman-Altman-2003} and \cite{PangScutari-IT-08}. In \cite{Iiduka-2010}, variational inequality theory is used to design a centralized power control algorithm for this scenario, in \cite{Altman-Altman-2003} the authors show that, if the assumption of S-modularity holds for the corresponding game, then best response dynamics \cite{Rose-CommMag-2011} converges to a generalized Nash equilibrium (GNE); in \cite{PangScutari-IT-08}, the authors provide, under the assumption of low interference, a sufficient condition for the convergence of the iterative water-filling algorithm to a GNE. It is worth noting that the works in \cite{Iiduka-2010}, \cite{Altman-Altman-2003} and \cite{PangScutari-IT-08} assume a compact and convex set of actions, i.e, the possible power allocation (PA) vectors may take any value in the corresponding simplex. 
\noindent
Conversely, in our work, we consider a finite action set by quantizing the possible available powers into a certain amount of levels. Basically, this is because in practice power levels must be expressed in a finite amount of bits. Moreover, several authors have pointed out that better global performance (e.g. spectral efficiency) is achieved when the set of PA vectors is substantially reduced \cite{Popescu-Rose-03}, \cite{Rose-Perlaza-2011}, \cite{Altman-BraessParadox}, \cite{Perlaza-Crowncom-09}. Indeed, this effect has been reported as a Braess kind paradox \cite{Braess-eng}. 
\noindent
In this paper, our contribution is twofold: first, we present a fully decentralized learning algorithm able to keep the SINR level above a certain threshold a high proportion of the time, by means of only \textit{one bit} feedback and relying only on local information \cite{Rose-CommMag-2011}; second, we analytically study the convergence properties and the convergence point, which is shown to be an efficient Nash equilibrium (NE) in terms of global performance.
\noindent
The paper is organized as follows. In Sec. \ref{System_Model}, we describe the wireless scenario and we formalize the problem; in Sec. \ref{Game_formulation}, we model the system as a game in normal form and in satisfaction form; in Sec. \ref{Trial_and_error} we present the trial and error algorithm as introduced in \cite{PaytonYoung-Trial-Error}; in Sec. \ref{Algo_properties}, we present a formal analytical study of the convergence properties (i.e., expected number of iterations to reach the NE and the satisfaction equilibrium (SE)), as well as, the expected fraction of time the system is at the NE and at the SE; in Sec. \ref{Numerical_results} we validate our analysis through numerical simulations; the paper is concluded in Sec. \ref{conclusion}.

\section{System model}\label{System_Model}
\begin{figure}
\centering
\def\svgwidth{0.7\columnwidth}
\begin{small}
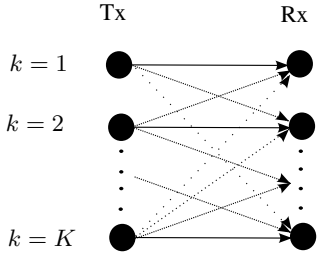
\end{small}
\caption{System model}\label{system_m}
\end{figure}

Let us consider the system described in Fig. \ref{system_m}. Here, a set $\mathcal{K} \defn \left\lbrace 1, ..., K \right\rbrace$ of transmitter-receiver pairs share a set  $\mathcal{C} \defn \left\lbrace b^{(1)}, ..., b^{(C)} \right\rbrace$ of orthogonal sub-bands. Transmitter $k$ is allowed to transmit over one sub-band at a time at a given power level. We denote by $p_k \in \mathcal{P}$, with $\mathcal{P} \defn \left\lbrace 0, ..., P_{MAX} \right\rbrace$, $\vert \mathcal{P} \vert = Q$, and $b_k \in \mathcal{C}$, the power level and the frequency sub-band chosen by transmitter $k$ respectively. 
\noindent
We denote by $\bs{p} = \left( p_1, p_2, ...,p_K \right)$ the network power allocation vector, by $\bs{b} = \left(b_1,b_2, ...,b_K \right)$ the spectrum occupation vector and by $\bs{a} = \left(a_1,a_2,...,a_K\right)$ a network configuration vector, where $a_k = (p_k,b_k)$.
To communicate, pairs have to achieve a sufficient SINR level, i.e., $SINR_k > \Gamma$ where we denote by $\Gamma$ the minimum SINR threshold allowing transmission. Receivers treat interference as Gaussian noise, thus:
\begin{equation} \label{sinr}
	SINR_k\left( \bs{a} \right) = \frac{p_k g_{k,k}^{(b_k)}} {\sigma^2 + \sum_{l \in \mathcal{K}\setminus k} p_l g_{k,l}^{(b_l)}  \ind{b_l = b_k}},
\end{equation}
where $g_{k,l}^{(b)}$ represents the channel power gain between transmitter $l$ and receiver $k$ over sub-band $b$, $\sigma^2$ is the power of the thermal noise assumed constant over the whole spectrum and $\ind{}$ represents the indicator function. In our scenario, we assume block-fading channels, i.e., channel realizations are time invariant for the whole transmission.
Our objective is the satisfaction of the SINR constraints for the largest possible set of pairs by using the lowest global energy consumption. Formally, we want the network  configuration vector $\bs{a}^*$ to be a solution of the following optimization problem
\begin{equation}
\label{Opt_problem}
\left\lbrace
\begin{array}{lr}
  \min_{\bs{p} \in \mathcal{P}^{K}} \sum_{k=1}^K p_k(n) & \\
	\mbox{s.t.  } SINR_k\left( \bs{a} \right) > \Gamma & \forall k \in \mathcal{K}^{*}
\end{array}			
\right. ,
\end{equation}
where we denote by $\mathcal{K}^{*} \subseteq \mathcal{K}$ the largest subset of links able to simultaneously achieve a sufficient SINR level. 
\noindent
Generally, to achieve this goal a central controller knowing all the network's parameters is required. In the following sections, we propose a decentralized algorithm demanding no information on the network which will 
 steer the system to a solution of \eqref{Opt_problem}. 
\section{Game Formulation}
\label{Game_formulation}
In this section, we model the scenario presented in Sec~\ref{system_m} in two different formulations: a normal-form game and a satisfaction-form \cite{Perlaza-JSTSP-2012} game. 
\subsection{Normal form formulation}
We model the network described above by the game in normal form
\begin{equation}
\mathcal{G}=\left( \mathcal{K}, \mathcal{A} , \left\lbrace u_k \right\rbrace_{k \in \mathcal{K}} \right).
\end{equation}
Here, $\mathcal{K}$ represents the set of players, $\mathcal{A}$ is the joint set of actions, that is,  $\mathcal{A} = \mathcal{A}_1 \times \mathcal{A}_2 \times ... \times \mathcal{A}_K$ where $\mathcal{A}_k = \mathcal{C} \times \mathcal{P}$ and we introduce the utility function $u_k: \mathcal{A} \rightarrow \mathds{R}$ defined by:
\begin{equation}\label{ut_function}
u_k(\bs{a}) = \frac{1}{1+ \beta}\left( \frac{P_{MAX} - p_k}{P_{MAX}} + \beta \ind{SINR_k(\bs{a}) >\Gamma} \right),
\end{equation}
where $\beta$ is a design parameter discussed in Sec \ref{Algo_properties}. This function has been designed to be monotonically decreasing with the power consumption, and monotonically increasing with the number of players who achieve the minimum SINR. In the following, we show that, with this utility function, the NE of the game $\mathcal{G}$ can solve the problem stated in \eqref{Opt_problem}. Moreover, note that to evaluate \eqref{ut_function} each transmitter only requires local information, since $\ind{SINR_k(\bs{a}) >\Gamma}$ can easily be fed back by the receiver with $1$ bit.
\noindent
\begin{definition}\emph{
(Interdependent game). $\mathcal{G}$ is said to be interdependent if for every non-empty subset $\mathcal{K}^+ \subset K$ and every action profile $\bs{a}=(\bs{a}_{\mathcal{K}^{+}},\bs{a}_{-\mathcal{K}^{+}})$\footnote{Here, $a_{-\mathcal{K}^+}$ refers to the action profile of all the players that are not in $\mathcal{K}^+$} such that $\bs{a}_{\mathcal{K}^{+}}$ is the action profile of all players in $\mathcal{K}^+$, it holds that:
\begin{equation}
\exists i \notin \mathcal{K}^{+},\exists \bs{a}'_{\mathcal{K}^{+}}\neq \bs{a}_{\mathcal{K}^{+}}: u_i(\bs{a}'_{\mathcal{K}^+},\bs{a}_{-\mathcal{K}^{+}}) \neq  u_i(\bs{a}_{\mathcal{K}^{+}},\bs{a}_{-\mathcal{K}^{+}})
\end{equation}
}
\end{definition}
In the following, we assume that game $\mathcal{G}$ is interdependent. This is a reasonable assumption, since, physically, this means that no link is isolated from the others. 
The solution concept used under this formulation is the Nash equilibrium, which we define as follows:
\begin{definition}\emph{
(Nash equilibrium in pure strategies). An action profile $\bs{a}^{*} \in \mathcal{A}$ is a NE of game $\mathcal{G}$ if $\forall$ $k \in \mathcal{K}$ and $\forall a'_k \in \mathcal{A}_k$
\begin{equation}
u_k(a^{*}_k,a^{*}_{-k}) \geq u_k(a'_k,a^{*}_{-k} ).
\end{equation}}
\end{definition}
To measure the efficiency of each NE, we introduce the \textit{social welfare} function, defined by the sum of all individual utilities: $W(\bs{a}) = \sum_{k=1}^K u_k(\bs{a})$.

\subsection{Satisfaction form} 
The satisfaction form is a game theoretical formulation modelling scenarios where players are exclusively interested in the satisfaction of their individual QoS constraints. Let us define the game as 
\begin{equation}
\mathcal{G}'=\left( \mathcal{K},\mathcal{A},\left\lbrace f_k \right\rbrace _{k \in \mathcal{K}} \right),
\end{equation}
where $\mathcal{K}$, $\mathcal{A}$ follow the previous definitions and the satisfaction correspondence $f_k: \mathcal{A}_{-k} \rightarrow \mathds{R}$ is defined by
\begin{equation}
f_k(\bs{a}_{-k})= \left( a_k \in \mathcal{A}_k : SINR_k{\left( a_k, \bs{a}_{-k},\right)} \geq \Gamma \right).
\end{equation}
\noindent
The solution concept used under this formulation is the satisfaction equilibrium (SE) defined as:
\begin{definition}
\label{SE_DEF}
\emph{
(Satisfaction equilibrium). A satisfaction equilibrium of game $\mathcal{G}'$ is an action profile $\bs{a'} \in \mathcal{A}$ such that $\forall k \in \mathcal{K}$, 
\begin{equation}
a'_k \in f_k \left( \bs{a}'_{-k} \right).
\end{equation}
}
\end{definition}
Moreover, we measure the effort of player $k$ due to the use of a particular action $\bs{a}_k$ by using the effort function \cite{Perlaza-JSTSP-2012} $\Phi_k: \mathcal{A}_k  \rightarrow \left[0,1\right]$. We can, then, define an efficient satisfaction equilibrium (ESE) as:
\begin{definition}\emph{
(Efficient satisfaction equilibrium). A satisfaction equilibrium $\bs{a}'$ is said to be \textit{efficient}, if $\forall k \in \mathcal{K}$
\begin{equation}
 a'_k \in \arg\min_{a_k \in f_k\left(\bs{a}'_{-k} \right)}\Phi_k(a_k)
\end{equation}}
\end{definition}
In brief, an ESE is an action profile where all players are satisfied and no player may decrease its individual effort by unilateral deviation.
\noindent
Since our optimization problem is to minimize the overall transmit power, we identify the effort by the function: $\Phi_k(a_k)=p_k$.
\section{Algorithm Description}
\label{Trial_and_error}
In this section, we briefly describe the trial and error (TE) algorithm introduced in \cite{PaytonYoung-Trial-Error}, \cite{pradelsky}. Later, we characterize the degrees of freedom of the system to fit in our scenario. In TE learning, each player $k$ locally implements a state machine, at each iteration $n$, a state is defined by the triplet:
\begin{equation}
\label{TE_state}
Z_k(n)= \left\lbrace m_k(n), \bar{a}_k(n), \bar{u}_k(n) \right\rbrace ,
\end{equation}
where $m_k(n) \in \left\lbrace C,C+,C-,D\right\rbrace$ represents a "mood", i.e., a characteristic that defines the machine reaction to its experience of the environment, $\bar{a}_k \in \mathcal{A}$ and $\bar{u}_k \in [0,1]$ represents a \textit{benchmark} action and \textit{benchmark} utility, respectively. There are four possible moods: content ($C$), watchful ($C-$), hopeful ($C+$), discontent ($D$). In the following, we characterize the behaviour of each player in every possible mood.
\begin{itemize}
\item {\textbf{Content}}
\end{itemize}
If at stage $n$ player $k$ is content, it uses the benchmarked action $\bak(n)$ with probability $(1-\epsilon)$ and experiments a new action $a'_k(n)$ with probability $\epsilon$. If at stage $n$ the player decided to experiment, at stage $(n+1)$ it evaluates the utility $u_{k}'(n+1)$ associated with $a'_k(n)$ as follows: if $u'_k(n+1) < \bar{u}_k(n)$ then $Z_k(n+1)=Z_k(n)$, otherwise if $u'_k(n+1) > \bar{u}_k(n)$, then, with probability $\epsilon ^ {G(u'_k(n+1)-\bar u_k(n))}$, a new action and utility benchmark are set out, i.e., $\bar u_k(n+1)=u'_k(n+1)$ and $\bar a_k(n+1)=a'_k(n)$, respectively. 
Here, $G(\cdot)$ must be such that:
\begin{equation}
\label{condition_on_G}
0 < G(\Delta u) < \frac{1}{2},
\end{equation}
we opt for a linear formulation: $G(\Delta u) = -0.2 \Delta u + 0.2$.
\begin{itemize}
\item \textbf{Hopeful-Watchful}
\end{itemize}
If player $k$ achieves an increment or a decrement in its utility without having experimented at the previous stage, then the mood become hopeful or watchful, according to the following rule: $(i)$ if $u'_k(n+1) > \bar u_k(n)$ then, $m_k(n+1) = C+$, $\bar a_k(n+1)=\bar a_k(n)$ and $\bar u_k(n+1)=\bar u_k(n)$; $(ii)$ if $m_k(n+1) = C-$, then $\bar a_k(n+1)=\bar a_k(n)$ and $\bar u_k(n+1)=\bar u_k(n)$.
\noindent
If player $k$ observes an improvement also at the next stage (i.e., $u'_k(n+2) > \bar u_k(n+1)$), then the mood switches to content and the benchmark utility is updated with the new one: $m_k(n+2)=C$ and $\bar u_k(n+2)= u'_k(n+1)$.
On the contrary, if a loss is observed also at the next stage (i.e., $u'_k(n+2) < \bar u_k(n+1)$), then the mood switches to discontent $m_k(n+2)=D$.
\begin{itemize}
\item \textbf{Discontent}
\end{itemize}
If player $k$ is discontent, it experiments a new action ($a'_k(n)$) at each step $n$. We refer to this behaviour as \textit{noisy} search. When the corresponding utility $u_k'(n+1)$ is observed, with probability $p=\epsilon^{F(u_k'(n+1)}$ the mood turns to content $m_k(n+1)=C$, a new action and utility benchmark are set up, $\bar{u}_k(n+1) = u_k'(n+1)$ and $\bar{a}_k(n+1) = a_k'(n+1)$, while, with probability $(1-p)$ it continues the noisy search. Note that function $F$ must be such that
\begin{equation}
\label{condition_on_F}
0 < F(u) < \frac{1}{2K},
\end{equation}
we opt for a linear formulation: $F(u) = -\frac{0.2}{K} u + \frac{0.2}{K}$.
%
\subsection{Algorithm properties}
Hereunder, we restate Theorem 1 in \cite{PaytonYoung-Trial-Error} and Theorem 1 in \cite{pradelsky} using our notation.
\begin{theorem}\emph{
\label{P-Y-th1}
Let $\mathcal{G}$ have at least one pure Nash equilibrium and let $\epsilon$ be small enough. Then, a pure Nash equilibrium is played at least $(1-\delta)$ of the time.}
\end{theorem}
This theorem introduces a different notion of convergence. Generally \cite{Rose-CommMag-2011}, we say that an algorithm converges when it approaches a certain solution as $n \rightarrow \infty$ while, here, it means that this solution is played with a high probability an high proportion of the total time.
\noindent
\begin{theorem}
\emph{
\label{P-Y-th2}
Let $\mathcal{G}$ have at least one pure Nash equilibrium and let each player employ TE, then a Nash equilibrium that maximizes the sum utility among all equilibrium states is played a large proportion of the time.}
\end{theorem}

Note that, generally, different equilibria are associated with different social welfare values. Learning algorithms available in the literature \cite{Rose-CommMag-2011}, \cite{Scutari-Algorithms-2008}, do not always take into consideration the problem of equilibrium selection, which is a central issue when aiming at global performance. 

\section{Main results}
\label{Algo_properties}
\subsection{Working point properties}
In this section, we present our results based on the previous analysis. Proofs are omitted due to space constraints. Based on the game theoretical formulation in Sec. \ref{Game_formulation} and the algorithm properties in Sec. \ref{Trial_and_error} we state the following:

\begin{theorem}
\emph{
Let $\mathcal{N} \neq \emptyset$ be the set of NE of $\mathcal{G}$, let $\beta > K$ and let us denote by $K_l$ the number of players satisfied at the $l$-th NE. Then, TE converges to the NE where $K_l$ is maximized.}
\end{theorem}
This theorem states that, if $\beta > K$, then TE converges to a state where the largest possible number of players are satisfied \textit{and} are at the NE. Here, $\beta$ represents the interest a network designer has in satisfying the largest set of players over the minimization of the network power consumption. The next two theorems allow us to link this result with the original global design problem expressed in \eqref{Opt_problem}. 
\begin{theorem}
\emph{
Let $(i)$ $\mathcal{A}^\dagger \neq \emptyset$ be the set of solutions of \eqref{Opt_problem} with $\mathcal{K}^* = \mathcal{K}$, $(ii)$ $\mathcal{N} \neq \emptyset$ be the set of NE of $\mathcal{G}$ and $\mathcal{N} \cap \mathcal{A}^\dagger \neq \emptyset$ and let $\beta > K$. Then, TE learning converges to an action profile $\bs{a}^*$ such that $\bs{a}^* \in \mathcal{N} \cap \mathcal{A}^\dagger$ and is an ESE.
}

\end{theorem}
This theorem links together the concept of ESE of game $\mathcal{G}'$, the NE of game $\mathcal{G}$ and the solutions of \eqref{Opt_problem}. Indeed, when the assumptions are met, the TE algorithm will reach a network state where: $(i)$ all players are satisfied, $(ii)$ the network power consumption is minimized. Note that, generally, it is possible for \eqref{Opt_problem} to have a solution that is not a NE of $\mathcal{G}$.
\begin{theorem}
\emph{
Let \eqref{Opt_problem} have no solution for $\mathcal{K}^* = \mathcal{K}$ and fix $\beta > K$. Let $K^*$ be the largest number of players that can be simultaneously satisfied and let  $\mathcal{K}^*_m$ be the $m$-th set, such that $\vert \mathcal{K}^*_m \vert = K^{*}$, where \eqref{Opt_problem} has a solution; let also be $\mathcal{A}^{*}_m$ the corresponding set of solutions. Let us define $\mathcal{A}^* = \bigcup_{m} \mathcal{A}^{*}_m$ and $\mathcal{N} \neq \emptyset$ the set of NE. Then, TE learning converges to an action profile $\bs{a}^*$ such that $\bs{a}^{*} \in \mathcal{N} \cap \mathcal{A}^{*}$.}
%
\end{theorem}
The previous theorem states that, when some players cannot satisfy their SINR condition, the TE algorithm selects the subset $\mathcal{K}^*_m$ among all possible $\mathcal{K}^*$ such that: $(i)$ the highest number of players are satisfied, $(ii)$ the network power consumption is minimized (with the unsatisfied players employing $0$ power).
\begin{corollary}
\emph{Let $\forall k$ and $\forall b$ be $g_{k,k}^{(b)} \geq \frac{\Gamma \sigma^2}{P_{MAX}}$, let $C \geq K$ and fix $\beta > K$. Then, TE converges to a solution of \eqref{Opt_problem}. }
\end{corollary}
Basically, this corollary means that, if transmitters and receivers are satisfiable on each channel (high SNR regime), then TE converges to an optimal working point.
\subsection{Convergence analysis}
\noindent
TE algorithm defines a discrete time Markov chain (DTMC) on the set of the states. Studying the behaviour of the algorithm on the complete chain is an intractable problem due to the number of states, transitions and parameters. In the following, we provide an approximated DTMC that allows us to estimate: $(a)$ the expected converging time at the NE and at the SE, $(b)$ the expected fraction of time the system is at the NE and at the SE.
\noindent
Under the light of the description made in Sec. \ref{Trial_and_error}, we state the following: $(i)$ the fraction of time spent in the \watchful or \hopeful states is negligible compared to the one in \discontent or \content one; $(ii)$ at any time, the probability of having more than one player \discontent is negligible.
 In the following, we assume $C>K$ and a simplified channel model, defined as
\begin{equation}
\left\lbrace
\begin{array}{lr}
  \label{chan_model} 
			 g_{k,k}^{(c)} = 1 & \forall k, \forall c \\
			 g_{j,k}^{(c)} = \frac{1}{2} &\forall k,\forall j \neq  k, \forall c 
\end{array}			.
\right.
\end{equation}
In Sec. \ref{Numerical_results} we will show that these results are good approximations also under less restrictive conditions.
\begin{figure}
\centering
\def\svgwidth{1\columnwidth}
\begin{small}
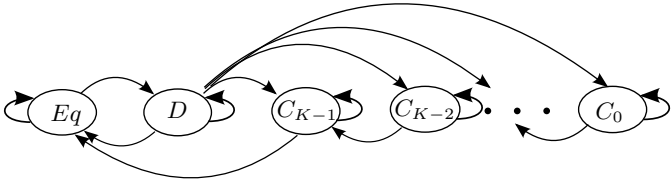
\end{small}
\caption{Markov chain describing the TE algorithm in the network. The state $Eq$ represents a state where all players are in equilibrium (i.e., SE or NE). $C_{K-k}$ represents a state where $K-k$ players are using a \textit{correct} action (i.e., an action that is satisfying or is optimal w.r.t. the others). D represents a state where one player is \textit{discontent}.}
\label{MC}
\end{figure}
\noindent
The resulting DTMC for studying TE behaviour is represented in Fig. \ref{MC}. When interested in convergence time and occupancy frequency of the NE, state $Eq$ represents the NE, and $C_{K-k}$ a state where $K-k$ players are using an individually optimal action and $D$ a state where one player is \textit{discontent}. 
The transition probabilities we evaluate are listed hereunder, the detailed description is omitted due to space constraints.


\begin{eqnarray}
	P(NE,D) =& \frac{K(K-1)^2\epsilon^2}{C^2} \left(\frac{Q-1}{Q}\right)^2 \label{NE->D}\\
	P(D,NE) =& \frac{\left(C-K+1)\right)}{CQ}\\
	P(D,C_{K-k}) =& \frac{\left(C-K+k\right)}{C^k}\frac{\left(K-1\right)!}{\left(K-k\right)!}\label{D->C_n}\\
	P \left(C_{K-k},C_{K-k-1}\right) =& (K-k)\frac{C-k}{CQ} \epsilon^{1+G\left( \Delta u \right)}.
\end{eqnarray}
\noindent
The analysis of this DTMC allows us to state the following theorems:
\begin{theorem}\emph{
The expected number of iterations needed before reaching the NE for the first time $\bar{T}_{NE}$ is bounded as follows:
\begin{eqnarray}
\bar{T}_{NE} &\leq& \frac{CQ}{\epsilon^{\left(1+G(\Delta u)\right)}\left(C-K\right)} \left(1 + \log \left(  \frac{ K \left(C-K+1\right) }{C+1}\right) \right)
\nonumber\\
\bar{T}_{NE} &\geq& \frac{CQ}{\epsilon^{\left(1+G(\Delta u)\right)}\left(C-K\right)}\left(\gamma + \log \left(  \frac{K \left(C-K\right) }{C}\right) \right);
\nonumber
\end{eqnarray}
where, $\gamma \simeq 0.577$ is the Euler-Mascheroni constant.}
\end{theorem}
Note that, the time demanded to converge is directly proportional to the degree of freedom (i.e., $\vert \mathcal{A}_k \vert = CQ$) and inversely to the experimentation probability $\epsilon$. Nonetheless, as we shall see, choosing a large $\epsilon$ increases the instability of the NE and, consequentially, the network performance. 
%
%
%

\begin{theorem}\emph{
The expected fraction of time the system is at a NE $(1-\delta)$ is:
\begin{equation}
(1-\delta) = \frac{1}{1+P(NE,D)T_{BNE}},
\end{equation}
where 
\begin{eqnarray}\nonumber
T_{BNE} &\leq& \sum_{k=1}^{K}  P(D,C_{K-k})T_{CNE}(k) + \frac{P(D,NE)}{\left(1-P(D,D)\right)^2}\\
\nonumber
T_{CNE}(k)&=& \frac{CQ}{\epsilon^{1+G\left(\Delta u\right)}\left( C-K \right)} \left(\gamma + \log \left(  \frac{ K \left(C-k+1\right) }{C+1}\right) \right)\\
	P(D,D) &=& 1-P(D,NE) - \sum_{k=1}^{K} P(D,C_{K-k}).\nonumber
\end{eqnarray}
}
\end{theorem}
Here, $(1-\delta)$ depends on $\frac{1}{\epsilon^2}$ as in \eqref{NE->D}. This means that, the larger the $\epsilon$ the shorter the time the system is at a NE.
\noindent
To evaluate convergence time and occupancy frequency of the SE we, again, make use of Fig. \ref{MC}. In this case, state $Eq$ represents the SE, $C_{K-k}$ is a state where $K-k$ players are satisfied and $D$ a state where one user is \textit{discontent}. The corresponding transition probabilities are listed hereunder.
\begin{eqnarray}
P(SE,D)=&\frac{K(K-1)^2\epsilon^2}{C^2}\left(\frac{Q-1}{Q}\right)^2\label{SE->D}\\
P(D,SE)=&\frac{\left(C-K+1\right)}{C}\\
P(D,C_{K-k})=&\frac{1}{C^k}\frac{\left(K-1\right)!}{\left(K-k\right)!} \left(C-K+k\right)\\
P \left(C_{K-k},C_{K-k-1}\right)=&(K-k)Q_{S}\frac{\left(C-k\right) \epsilon^{1+G\left( \Delta u \right)}}{CQ}.\label{SE_C_K_eq}
\end{eqnarray}
Given the model in \eqref{chan_model} and $C>K$, the term $Q_{S} < Q$ represents the number of power quantization levels that a player can employ to successfully achieve $SINR_k > \Gamma$ on any free channel.
\noindent
We can, then, state the following theorems:
\begin{theorem}
\emph{The expected number of iterations needed before reaching the SE for the first time $\bar{T}_{SE}$ is bounded as follows:
\begin{eqnarray}\nonumber
\bar{T}_{SE} &\leq & \frac{CQ/Q_{S}}{\epsilon^{\left(1+G(\Delta u\right)}\left(C-K\right)} \left(1 + \log \left(  \frac{ K \left(C-K+1\right) }{C+1}\right) \right)\\
\bar{T}_{SE} &\geq & \frac{CQ/Q_{S}}{\epsilon^{\left(1+G(\Delta u\right)}\left(C-K\right)} \left(\gamma + \log \left(  \frac{K \left(C-K\right) }{C}\right) \right) \nonumber .
\nonumber
\end{eqnarray}
}
\end{theorem}
Under assumption \eqref{chan_model}, being satisfied is a weaker condition than being at the NE, thus, it results that $T_{NE} \geq T_{SE}$. Predictably, larger $P_{MAX}$ and lower $\Gamma$ increasing $Q_S$, are able to improve the converging speed.
\begin{theorem}
\emph{
The expected fraction of time the system is at a SE $F_{SE}$ is:
\begin{equation}
F_{SE}=\frac{1}{1+P(SE,D)T_{BSE}}
\end{equation}
where
\begin{eqnarray}\nonumber
T_{BSE} &\leq &\sum_{k=1}^{K}  P(D,C_{K-k})T_{CSE}(k) + \frac{P(D,SE)}{\left(1-P(D,D)\right)^2}\\
\nonumber
T_{CSE}(k)&= &\frac{CQ}{\epsilon\left( C-K \right) Q_{S}}\left(\gamma + \log \left(  \frac{ K \left(C-k+1\right) }{C+1}\right) \right) \\
P(D,D) &=& 1-P(D,SE) - \sum_{k=1}^{K} P(D,C_{K-k})\nonumber .
\end{eqnarray}
}
\end{theorem}
\section{Simulation results}
\label{Numerical_results}
\begin{figure}
\centering
\def\svgwidth{0.2\columnwidth}
\vspace{0cm}
\includegraphics[width=0.52\textwidth]{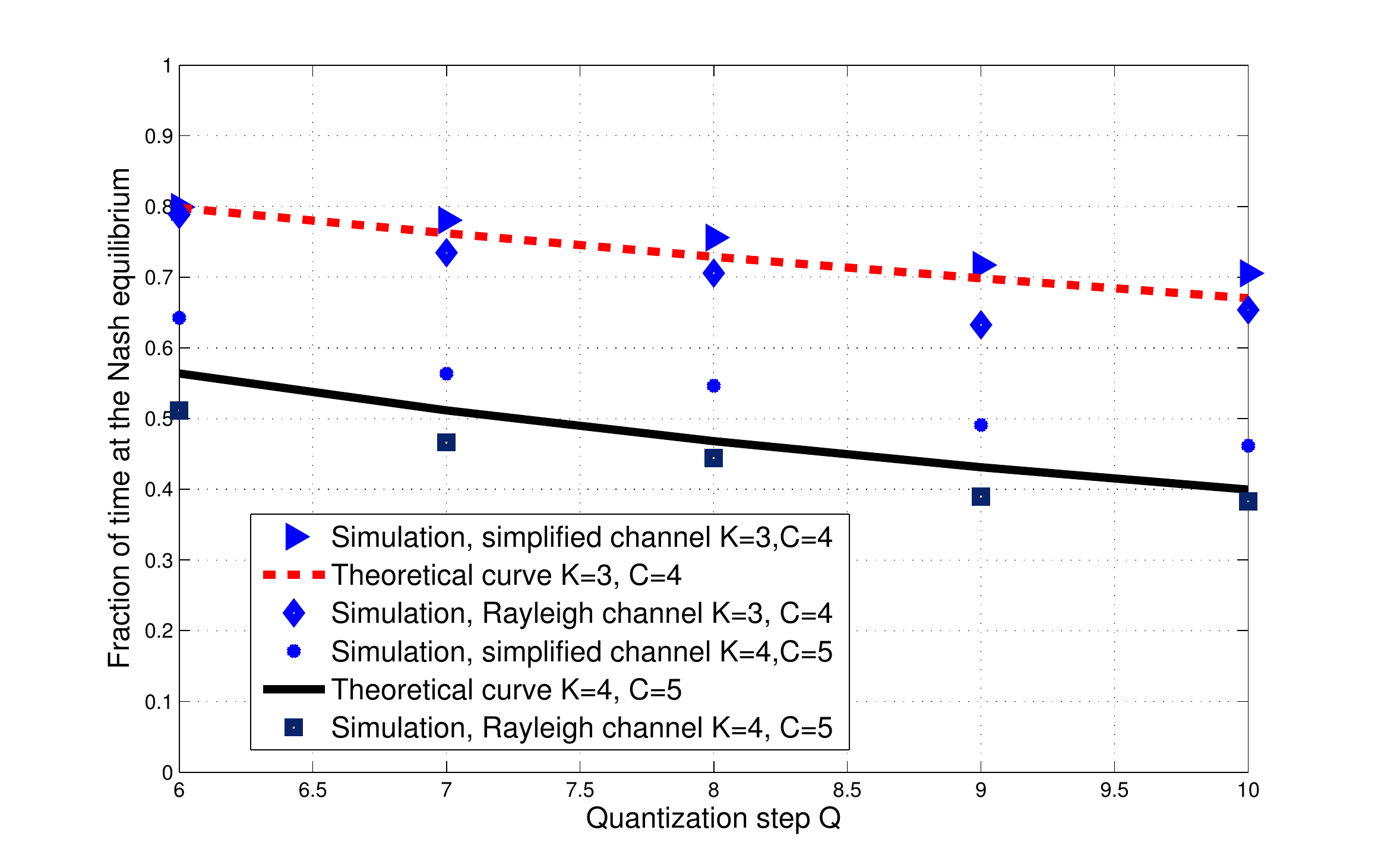}
\vspace{0cm}
\caption{Fraction of time the system is at the NE. Comparison between theoretical line and simulation results for two set of data and different channels: Rayleigh and simple one as in \eqref{chan_model}.}
\label{NE_frac}
\end{figure}

\begin{figure}
\centering
\def\svgwidth{0.2\columnwidth}
\vspace{0cm}
\includegraphics[width=0.52\textwidth]{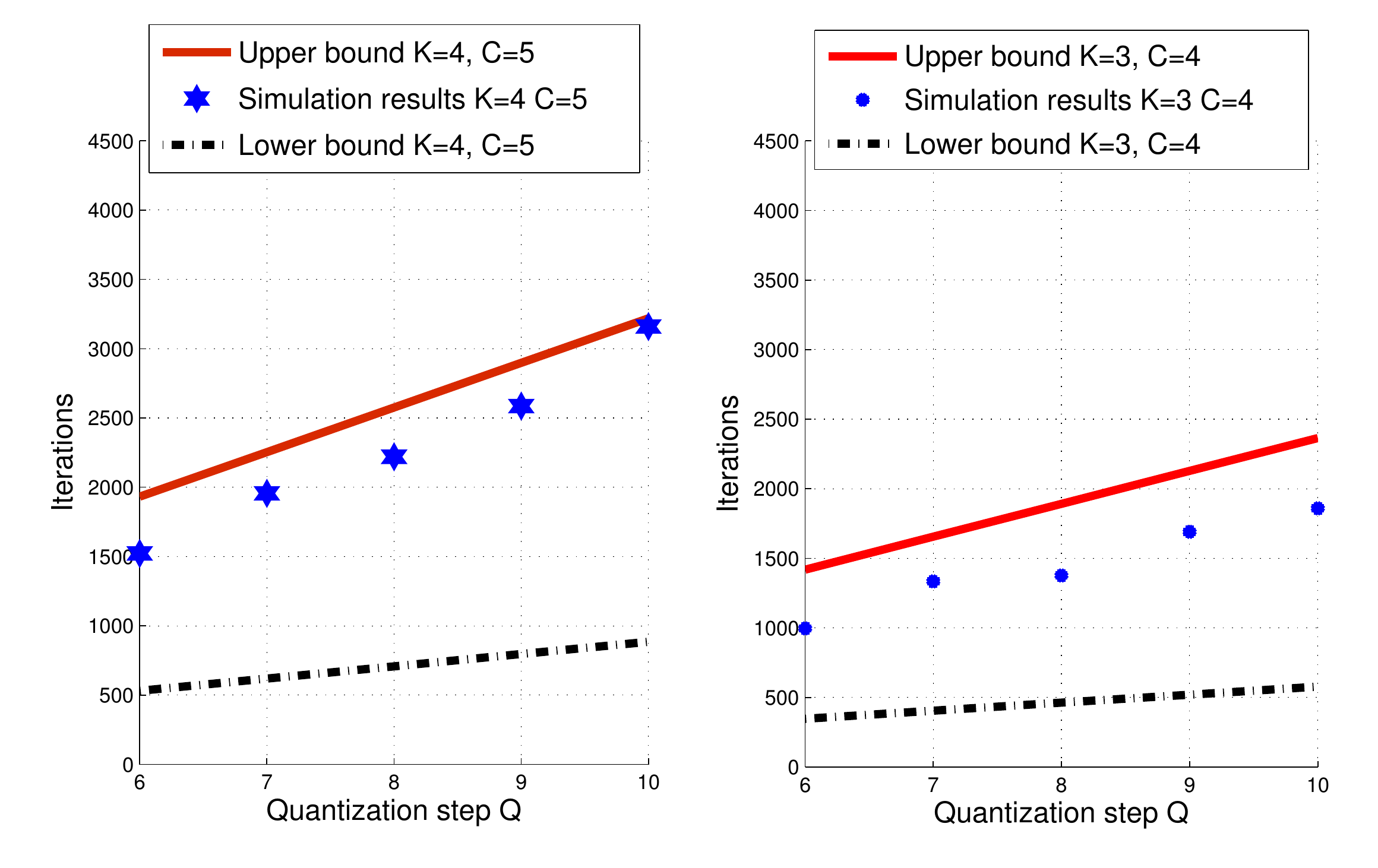}
\vspace{0cm}
\caption{Expected time for the system to reach the Nash equilibrium. On the left, results for a system with $K=4$ players and $C=5$ channel. On the right, results for a system with $K=3$ players and $C=4$ channels.}
\label{rising_time}
\end{figure}

\begin{figure}
\centering
\def\svgwidth{0.2\columnwidth}
\vspace{0cm}
\includegraphics[width=0.52\textwidth]{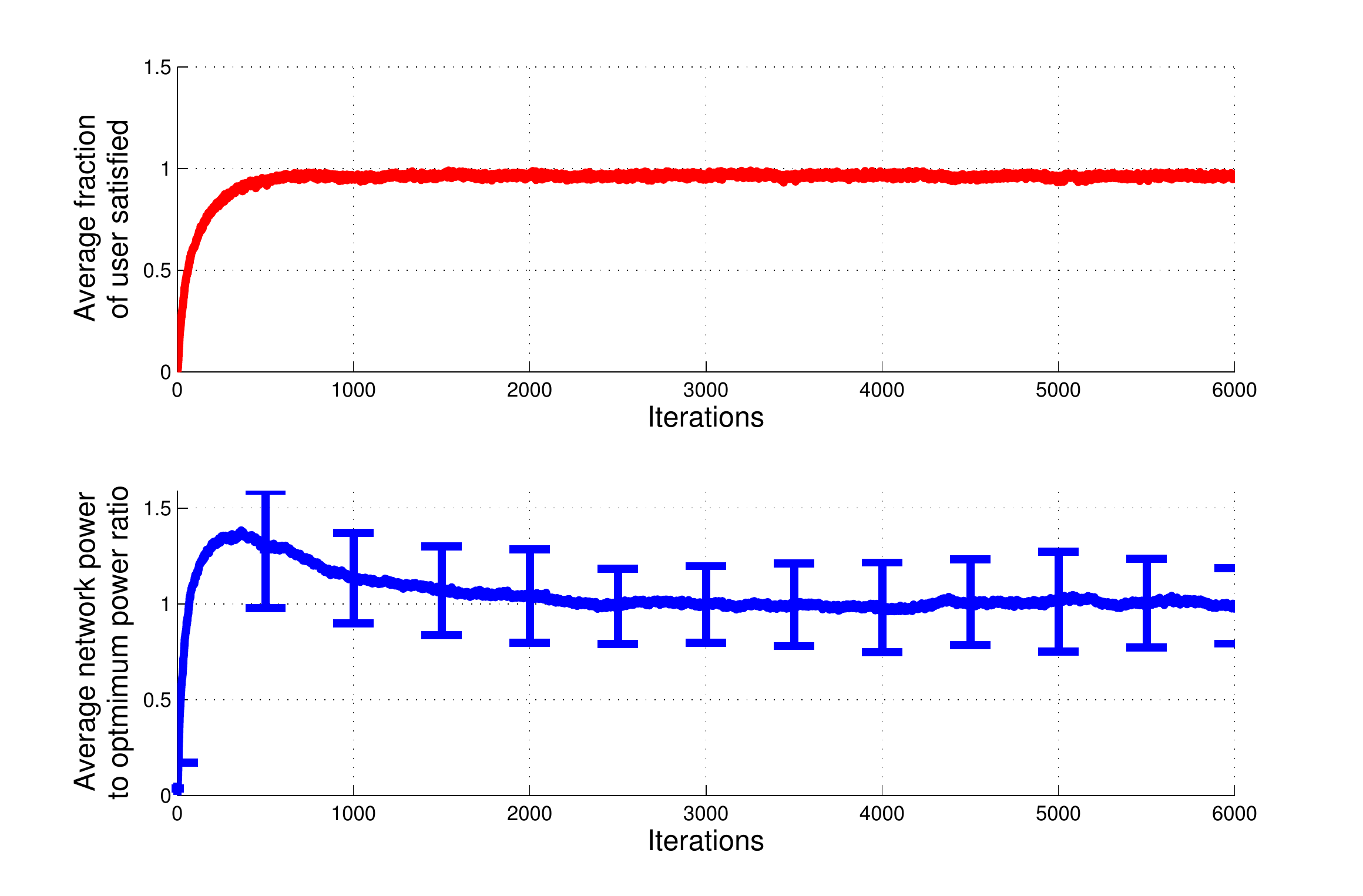}
\vspace{0cm}
\caption{Average fraction of players satisfied versus iterations, and ratio between the average power employed and the optimal level of power that should be employed to satisfy the whole network. Available number of channel $C=5$, number of players $K=4$, quantization level $Q=8$.}
\label{FIG_sat_vs_power}
\end{figure}

The purpose of this section is threefold. First, we run simulations to numerically validate the DTMCs introduced in Sec. \ref{Algo_properties}, second, we validate the results on more general channel models, then, we evaluate the performance of the algorithm in terms of satisfaction and power employed. 
The first two experiments have been run for two different sets of parameters. The first set is composed by: $K=3$, $C=4$, $\epsilon=0.02$ and $6 \leq Q \leq 10$. The second set is composed by $K=4$, $C=5$, $\epsilon=0.02$ and $6 \leq Q \leq 10$. In our first experiment, we run $10^7$ iterations to estimate $(1-\delta)$ under two different channel models: the simple channels expressed in \eqref{chan_model} and a Rayleigh channel. The results are summarized in Figure \ref{NE_frac}. As we can see, the analysis, brought on particular channel model, proves to be sufficiently precise also under more general formulations.
\noindent
In our second experiment, we estimated the converging time and compared with the analytical results in Figure \ref{rising_time}. As we can see, increasing the action set dimension, i.e., increasing $C$ or $Q$, brings slower convergence rate since the algorithm requires more time to explore all the possibilities. 
Note that, here, convergence time means the time needed by the system to work at the NE for the first time. The third experiment's parameter set is composed as follows: $K=4$, $C=5$, $\epsilon=0.02$ $Q=8$ with the simplified channel model as in \eqref{chan_model}. Here, we have run $10^3$ tests, each one composed by $6000$ iteration of TE. The results are showed in Fig \ref{FIG_sat_vs_power}, where the upper curve represents the fraction of players satisfied, while the lower curve represents the ratio between the average power employed by the network and the optimal power that should be employed to satisfy all the players. In average, in accordance with Figure \ref{NE_frac}, the system reaches an optimal equilibrium (all players satisfied and minimum amount of power employed) after around $2200$ iterations. Note that, even though, for some specific scenarios, this number may be too high, the configurations selected by the algorithm before the convergence are just slightly inefficient. Indeed, a configuration where all the players are able to satisfy their SINR constraints is averagely reached after $600$ iterations. Moreover, before this, we observe that only a fraction of satisfiable players is satisfied, in spite of the amount of power used.

\section{conclusion}
\label{conclusion}
In this paper, we have studied a power control problem in a self configuring decentralized network. We presented a new decentralized algorithm 
able to steer the network into a working point where the maximum number of transmitter-receiver pairs achieves a sufficient SINR while minimizing the network power consumption. The algorithm does not assume any prior knowledge of the network and can learn an efficient equilibrium with only \textit{one} bit of feedback. By assuming a particular channel realization, we have analytically estimated the expected performance of the algorithm through a Markov chain description of the algorithm behaviour. Finally we have shown through Monte-Carlo simulations that the analysis is approximatively correct also for general channel models. 
\balance
\section{Acknowledgement}
This research work was carried out in the framework of the CORASMA – EDA Project B-0781-IAP4-GC.
\bibliographystyle{IEEEtran}
\bibliography{GT}

\end{document}

%% file: scenario1.pdf_tex

\begingroup
  \makeatletter
  \providecommand\color[2][]{%
    \errmessage{(Inkscape) Color is used for the text in Inkscape, but the package 'color.sty' is not loaded}
    \renewcommand\color[2][]{}%
  }
  \providecommand\transparent[1]{%
    \errmessage{(Inkscape) Transparency is used (non-zero) for the text in Inkscape, but the package 'transparent.sty' is not loaded}
    \renewcommand\transparent[1]{}%
  }
  \providecommand\rotatebox[2]{#2}
  \ifx\svgwidth\undefined
    \setlength{\unitlength}{455.28585597pt}
  \else
    \setlength{\unitlength}{\svgwidth}
  \fi
  \global\let\svgwidth\undefined
  \makeatother
  \begin{picture}(1,0.52382609)%
    \put(0,0){\includegraphics[width=\unitlength]{scenario1.pdf}}%
    \put(0.19997205,0.5305116){\color[rgb]{0,0,0}\makebox(0,0)[lt]{\begin{minipage}{0.13026319\unitlength}\raggedright Tx\end{minipage}}}%
    \put(0.58759025,0.52679023){\color[rgb]{0,0,0}\makebox(0,0)[lt]{\begin{minipage}{0.09537126\unitlength}\raggedright Rx\end{minipage}}}%
    \put(0.00621957,0.38785316){\color[rgb]{0,0,0}\makebox(0,0)[lb]{\smash{$k=1$}}}%
    \put(-0.00156308,0.2587646){\color[rgb]{0,0,0}\makebox(0,0)[lb]{\smash{$k=2$}}}%
    \put(0.00243326,0.01099192){\color[rgb]{0,0,0}\makebox(0,0)[lb]{\smash{$k=K$}}}%
  \end{picture}%
\endgroup

%% file: MC_EQ.pdf_tex

\begingroup
  \makeatletter
  \providecommand\color[2][]{%
    \errmessage{(Inkscape) Color is used for the text in Inkscape, but the package 'color.sty' is not loaded}
    \renewcommand\color[2][]{}%
  }
  \providecommand\transparent[1]{%
    \errmessage{(Inkscape) Transparency is used (non-zero) for the text in Inkscape, but the package 'transparent.sty' is not loaded}
    \renewcommand\transparent[1]{}%
  }
  \providecommand\rotatebox[2]{#2}
  \ifx\svgwidth\undefined
    \setlength{\unitlength}{403.7610064pt}
  \else
    \setlength{\unitlength}{\svgwidth}
  \fi
  \global\let\svgwidth\undefined
  \makeatother
  \begin{picture}(1,0.26327456)%
    \put(0,0){\includegraphics[width=\unitlength]{MC_EQ.pdf}}%
    \put(0.24445854,0.11486356){\color[rgb]{0,0,0}\makebox(0,0)[lt]{\begin{minipage}{0.06867257\unitlength}\raggedright $D$\end{minipage}}}%
    \put(0.4094799,0.1164357){\color[rgb]{0,0,0}\makebox(0,0)[lt]{\begin{minipage}{0.05897368\unitlength}\raggedright $C_{K-1}$\end{minipage}}}%
    \put(0.58612544,0.11826113){\color[rgb]{0,0,0}\makebox(0,0)[lt]{\begin{minipage}{0.05449949\unitlength}\raggedright $C_{K-2}$\end{minipage}}}%
    \put(0.88788687,0.1132208){\color[rgb]{0,0,0}\makebox(0,0)[lt]{\begin{minipage}{0.07090485\unitlength}\raggedright $C_{0}$\end{minipage}}}%
    \put(0.06993182,0.11062067){\color[rgb]{0,0,0}\makebox(0,0)[lt]{\begin{minipage}{0.12177729\unitlength}\raggedright $Eq$\end{minipage}}}%
  \end{picture}%
\endgroup